\documentstyle[12pt]{article} \textwidth = 15 truecm \textheight = 21
truecm 
\hoffset = - 0.6 truecm \voffset=-1.5 truecm

\def\lsim{\raise0.3ex\hbox{$<$\kern-0.75em\raise-1.1ex\hbox{$\sim$}}}
\def\gsim{\raise0.3ex\hbox{$>$\kern-0.75em\raise-1.1ex\hbox{$\sim$}}}

\begin{document} \begin{center} {\large \bf Reply to K A Kirkpatrick}\vskip 5
truemm {\bf B. d'Espagnat} \vskip 3 truemm

{\it Laboratoire de Physique Th\'eorique, UMR 8627 CNRS,\\ Universit\'e
Paris XI, B\^atiment 210, 91405 Orsay Cedex, France,\\ E-mail :
Espagnat@th.u-psud.fr} \end{center}

\begin{abstract}   
This is a reply to an article with the same title in which
Kirkpatrick claimed that the considerations I put forward some thirty
years ago on quantum mixtures are incorrect. It is shown here that
Kirkpatrick's reasoning is erroneous.
\end{abstract}

\baselineskip=18pt

\noindent {\bf 1. Introduction and a preliminary remark.} \\

In a recent paper [1] K. A. Kirkpatrick criticised the distinction I
introduced long ago [2] (see also [3] and [4]) between the notions of
proper and improper mixtures. In the present article I shall explain
why, in my view, his criticism is unfounded. The subject must first be
introduced and, quite appropriately, Kirkpatrick did this by referring
to von Neumann's book [5]. ``Von Neumann  -- he wrote -- introduced
mixtures of pure ensembles into quantum mechanics exactly in the manner
of classical probability, as a matter of ignorance. Introducing the
statistical operator (density matrix) as the descriptor of a mixture he
said: ``if we do not even know what state is actually present -- for
example, when several states $\phi_1$, $\phi_2$, ... with respective probabilities
$w_1$, $w_2$, ... constitute the description -- then the statistical operator is $\rho
= \Sigma_s w_s|\phi_s><\phi_s|$''. Besides, Kirkpatrick also made a recall: ``... von
Neumann proved -- he wrote -- that the unique statistical descriptor of a
subsystem $S$ of a joint system $S+M$ is given by the partial trace $\rho^S =
Tr_M\{\rho^{S+M}\}$. [...] This statistical operator can always be expressed (in
many ways) as a convex sum of pure-state projectors, exactly in the
form of the `ignorance' mixture first introduced''. In his introduction
Kirkpatrick then further noted that I choosed to call proper mixtures
the mixtures defined by von Neumann in the first stated manner,
improper mixtures those made up of the subsystems $S$ of a composite
system $S+M$, and that I somehow questioned the possibility of
identifying the two notions (this is why I introduced these epithets). \par

Needless to say: on all this I fully agree. Where I first start
disagreeing to some extent with Kirkpatrick is on his precise
characterizing of my questioning. He wrote: ``D'Espagnat claims that an
ignorance interpretation of the improper mixture is mathematically
inconsistent''. This formulation does not fully satisfy me in that it
conveys the - wrong - idea that the reason why I claim the improper
mixture cannot be given an ignorance interpretation is essentially of a
mathematical nature. This, actually, is not the case. Basically, the
reason in question is, as we shall see, not mathematical but
logico-semantical. It consists of the fact that, in science, all the
words we use must be have a meaning, so that, when we say that a given
mathematical description is consistent with such and such an
interpretation we must be able to explain the meaning of the words by
means of which the said interpretation is stated. My claim essentially
is that, when this is done concerning the word ``ignorance'', then, the
``improper mixtures'' represented by the partial traces of a composite
system pure state statistical operator cannot be given an ignorance
interpretation. It is therefore this claim that my critics should aim
at disproving. We shall see that Kirkpatrick's paper does not succeed
in doing so. \par

In the next section we shall go, properly speaking, into
this matter. Before that, however, it is appropriate that a preliminary
remark should be made, concerning a notion Kirkpatrick makes use of. \\

\noindent {\it Preliminary remark}. \par

It has to do with the ``rule of distinguishablity''.
By this name Kirkpatrick referred to the well-known rule of quantum
computation: ``when the alternative processes are indistinguishable,
square the sum of their amplitudes, when distinguishable, sum the
squares of their amplitudes'', a rule Feynman frequently used in his
books. It is true of course that, as a guide and as a ``short cut''
sparing tedious computations, the rule in question is a most useful
one. It should however be well noted that it does not rank among the
basic quantum mechanical rules or ``axioms'' (such as the correspondence
between observables and self-adjoint operators, the quantum law of
evolution or the generalized Born rule concerning probabilities of
observations). Actually, it is merely a consequence of the latter. It
simply follows from the fact that, whenever a system $S$ interacts with a
system $M$ that may react to this impact in a nonnegligible way, in order
to study what happens to $S$ it is necessary to consider the wave
function of the composite system $S+M$. If we are interested in $S$ alone
we must of course sum over all the conceivable measurement results
concerning $M$ (hence, technically, ``trace $M$ out''), and this normally
leads to the disappearance of cross-terms involving amplitudes
concerning $S$. The distinguishability rule follows. This shows, first
that within the realm of questions concerning which the rule of
distinguishability is applicable its use is in principle redundant (in
last resort, what really counts is the detailed experimental
arrangement and it is on its basis that one must argue in case of
doubt) and second,  more importantly, that its range of applicability
is well defined. In fact, it is limited to the type of computational
problems for the investigation of which it was conceived and proved.
This means, the rule is fully reliable when we want to derive, from our
knowledge of how a system was prepared, the probabilities we have of
getting such and such measurement outcomes. But we would have no right
to extend it to the conceptual analysis of basic questions falling,
partly or totally, outside this range. Trying to apply the said rule to
such questions may amount to depriving oneself of any possibility of
defining notions that are needed for their very formulation. And
indeed,  in the next section it will become apparent that this is
precisely the kind of error Kirkpatrick fell into. \\

\noindent {\bf 2. On Hugues'argument.} \\

 We can now turn to our subject proper.
Kirkpatrick's discussion of my standpoint starts (his Section 3) as
follows: ``D'Espagnat insists that the improper mixture, although
represented by the same statistical operator as the proper mixture,
does not represent a mixture of ensembles in pure states $\{|\phi_s>\}$; the
ignorance interpretation may not be applied to it''. He then tries to
show that this view is flawed. \par

To this end, he first considers the
argument - very similar to my own - by means of which R.I.G.Hughes [6]
justified this distinction; so, let us begin by analysing what he
objects to Hugues. According to Kirkpatrick, Hugues' argument runs as
follows. ``Consider a composite system $S+M$ in the pure state $\rho$, of which
the component states are the mixed states $\rho^S$ and $\rho^M$. For the sake of
the argument assume that $\rho^S = a_1|u_1><u_1| + a_2|u_2><u_2|$, while $\rho^M =
b_1|v_1><v_1| + b_2|v_2><v_2|$, with $a_1 \not= a_2$ and $b_1 \not= b_2$ so there are no
problems of degeneracy. Then, according to the ignorance interpretation
of $\rho^S$ and $\rho^M$, system $S$ is really in one of the pure states $|u_1>$ or $|u_2>$
and system $M$ is really in one of the pure states $|v_1>$ or $|v_2>$. But this
would mean that the composite system is really in one of the four
states $|u_j>|v_k>$, with probabilities $a_ib_k$ respectively - in other words,
that the composite system is in a mixed state. Since this contradicts
our original assumption, the ignorance interpretation simply will not
do''. \par

After having thus, for the benefit of the discussion, reproduced
Hugues' reasoning, Kirkpatrick claimed that it is wrong. ``This argument
is so clearly stated - he wrote - that its error stands out''. And he
explained: ``the claim that ``the composite system is in a mixed state''
is not supportable - nothing external to $S+M$ distinguishes those states
$|u_i>|v_k>$ from one another. We must add the state vectors (not the
projectors) $|\Psi> = \Sigma_{j,k}\psi_{jk}|u_jv_k>$ - a pure state''. \par

Sweeping as these
statements are, I claim they are in fact unjustified and incorrect.
Their first defect is that Kirkpatrick's reference to
non-distinguishability is out of place. The problem we are here faced
with is not one of calculating the probabilities we have of observing
this or that on a system $S$, given the way $S$ was prepared. It is a
general problem of interpretating the formalism. The arguments
developed in Section 1 above (Preliminary Remark) show that the idea or
trying to apply the rule of distinguishability to such questions is
unjustified. \par

The second defect in Kirkpatrick's rebuttal of Hugues'
argument can be described as follows. Note first that, in it,
Kirkpatrick  tacitly endorses Hugues' assertion that the considered
composite system is really in [but] one of the four states $|u_j>|v_k>$.
Indeed, in view of the starting assumption he accepted - that $S$ is
really in one of the pure states $|u_1>$ or $|u_2>$ and $M$ is really in one of
the pure states $|v_1>$ or $|v_2>$ -- he could not reject the said assertion
(which, incidentally, means that in an ensemble of such composite
systems some of them are in state $|u_1>|v_1>$, some others in state
$|u_1>|v_2>$ etc.). But on the other hand - and this is precisely the point
that makes his reasoning inconsistent - while he seems oblivious of
what this assertion usually means (in terms of differences in possessed
values), he offers no alternative definition of its meaning. Indeed,
nowhere does he state what the words ``is really'' actually mean to him.
However, as already stressed, in any statement aiming at objectivity
all the words used should have a meaning, so that, when we claim that a
given interpretation is consistent we must be able to explain what the
words expressing it actually signify. \par

Now, in science a statement is
meaningful only if it corresponds, directly or indirectly, to some
conceivable piece of experience. Consequently, when Kirkpatrick,
following Hugues, states that the composite system is really in one of
the states $|u_j>|v_k>$ he should be able to explain what he means by
referring to some possible experience. Not necessarily to an experience
we actually have, but at least to one that, conceivably, some people
could have. This, however, he did not do. And if we ourselves try to
fill up this gap, we find that there is but one possibility. It
consists in identifying the expression ``System $S$ is really in state
$|u>$'' to the conditional statement: ``if, on $S$, somebody measured an
observable $G$ having $|u>$ among its eigenvectors he/she would, with
certainty (probability 1), get as an outcome the precise eigenvalue of
which $|u>$ is an eigenvector. \par

To see what this implies let us focus on
the simplest possible case (implicitly but appropriately used by both
Hugues and Kirkpatrick for introducing the problem), namely the one in
which both the Hilbert spaces $H^S$ and $H^M$ of systems $S$ and $M$ are
two-dimensional. Let then $G$ be the observable of $S$ (a spin component
for example) that has $\{|u_1>,|u_2>\}$ as eigenvectors and let $g_1$ and $g_2$ be
the corresponding eigenvalues. Similarly, let $R$ be the observable of $M$
that has $\{|v_1>,|v_2>\}$ as eigenvectors and let $r_1$ and $r_2$ be the
corresponding eigenvalues. Hugues and Kirkpatrick both assert that
system $S$ is really in one of the two states $|u_j>$ and that the composite
system $S+M$ is really in one of the four $|u_j>|v_k>$ states of $S+M$. So,
according to them, if we consider an ensemble $\widehat{E}$ of such $S+M$ systems the
latter must be distributed in four subensembles labelled $i,k$ ($i,k = 1$
or 2). Let us then consider the subensemble $\widehat{E}^{i,n}$ of $\widehat{E}$ labelled $j=i$ and
$k=n$. In it, all the systems $S$ are really in state $|u_i>$. From the
foregoing definition, we thereore know that if $G$ were measured on an
element of this subensemble there is a probability 1 that the outcome
$g_i$ would be obtained. Then, however, we may resort to a well known (and
easily proved!) lemma that, partly using the above defined notations,
can be stated as follows. Let $Q$ be an ensemble of composite systems
$S+M$, let, again, $G$ be an observable nondegenerate in $H^S$, having
$\{|u_1>,|u_2>\}$ as eigenvectors and $g_1$, $g_2$ as the corresponding eigenvalues
and let the probability be 1 that, within $Q$, the outcome of the
measurement of $G$ be $g_i$. Then, the statistical operator (density matrix)
describing $Q$ factorizes, with $|u_i><u_i|$ as a factor. Here this lemma
applies, so that we know that the statistical operator $\rho^{i,n}$ describing
$\widehat{E}^{i,n}$ has $|u_i><u_i|$ as a factor. For the same reason we know that it has
$|v_n><v_n|$ as a factor. It therefore is:

\begin{equation}
\rho^{i,n} = |u_iv_n><u_iv_n| \quad .					
\end{equation}

Then, however, nobody can deny that, $a_j$ and $b_k$ being the proportions
defined by Hugues in his example, $\widehat{E}$ is describable by the density
matrix

\begin{equation}
\rho ' = \Sigma_{j,k} a_jb_k \ \rho_{j,k}
\end{equation}					

\noindent for indeed, from the very way in which this $\widehat{E}$ has been constructed it
follows that the probabilities concerning the outcomes of any
measurements whatsoever that we could choose to perform on its elements
are obtained by first evaluating the corresponding probabilities $p_{j,k}$
on each $\widehat{E}_{j,k}$ and then combining them according to the usual laws of
combined probabilities, in the form

\begin{equation}
\Sigma_{j,k} a_j b_k \ p_{j,k} \ ;
\end{equation}						

\noindent and it has been common knowledge ever since the appearance of von
Neumann's book that these probabilities are exactly those yielded by
$\rho '$.\par

But then, whether Kirkpatrick likes it or not, $\widehat{E}$ is quite obviously not
describable as a pure case $|\Psi > = \sum_{jk} \psi_{jk}|u_iv_k>$ since $\rho '$ is not a
projector ($\rho^{'2} \not= \rho '$), a fact implying in particular that, in whatever
way we decide to choose the coefficients $\psi_{jk}$, there are observables
concerning which $\rho '$ yields verifiable predictions differing from those
yielded by the pure case $\rho = |\Psi><\Psi|$.\par

It follows from this that - again, however we choose the $\psi_{jk}$  -- the
ensemble of the $S$ systems whose density matrix is obtained by partial
tracing of this $\rho$ over the Hilbert space of $M$ and the ensemble of the $M$
systems obtained by the symmetrical procedure (exchanging symbols $S$ and
$M$) cannot be mixtures defined, a la von Neumann, as a matter of
ignorance - that is, by combining subensembles endowed with different
characteristics (``proper mixtures'' in my terminology) - since, as we
just showed, if they were, the ensemble of the composite $S+M$ systems
would be describable by $\rho '$, which it is not. To sum up, we here have
been careful to give a meaning to the expression ``is really'', used by
Hugues and endorsed by Kirkpatrick, and consequently also to the
expression ``ignorance interpretation'', similarly used by these authors,
and this has led us to agree with Hugues  - and disagree with
Kirkpatrick - in asserting that the ignorance interpretation cannot be
applied to the ensembles yielded by the just described partial tracing
operation. \\

\noindent {\it Remark} \par \nobreak

 The foregoing argument may appear somewhat roundabout since we
might consider that, as soon as Kirkpatrick granted that the composite
system ``is really'' in one of the four states $|u_jv_k>$, he was thereby
forced to admit that it is describable by (2) and cannot, therefore, be
in a pure state. I myself tend to be convinced by this simplified
argument but the very existence of Kirkpatrick's paper shows that this
standpoint is not shared by everybody. The reason may be that, so long
as one just ponders on formulas without making precise what words mean
-- by referring to experience --, some vagueness remains that leaves a
place for disputable views. So, after all, there {\it is} a reason for
considering that the argument above is not totally redundant. \\

\noindent {\bf 3. The peculiarity of Quantum Mechanics.} \\

The interpretation of Quantum
Mechanics always raised conceptual problems. It is natural that
questions concerning the basic nature of quantum mixtures should not be
totally independent from these problems and it is therefore appropriate
that we should here have a look at the latter. \par

For that purpose, let us
make a detour to classical physics. It may be considered that - perhaps
setting apart classical statistical mechanics which is a debatable case
- classical physics, considered as a universal theory, was
ontologically interpretable. This does not mean that such an
interpretation was logically necessary. It was not. But it does mean
that it was admissible, in the sense that it did not generate
contradictions. All the fields and particles that appeared in the
classical formulas could without difficulty be viewed as being really
existing entities so that, when their values or, respectively,
positions were measured, the outcomes of the measurements could,
without qualms, be interpreted as revealing the values these quantities
actually had. I use to express this fact by saying that the
corresponding statements were ``strongly objective'' ones. As is well
known, the same does not hold true in quantum mechanics where, for
example, interpreting the wave function as being a real entity in the
above sense leads to a host of conceptual difficulties (nature of
collapse and so on). To be sure, quite a number of physicists still go
on thinking that all this questioning is just an old story. That such
interpretational problems were satisfactorily solved by Bohr a long
time ago. In a sense they are right. Bohr could write with confidence:
``The description of atomic phenomena has [É] a perfectly objective
character'' [7]. But let us have a look at the remaining part of this
sentence of him. It reads: ``... in the sense that no explicit reference
is made to any individual observer and that therefore [...] no ambiguity
is involved in the communication of observation'' (emphasis ours). In
other words, according to Bohr atomic physics is indeed objective, but
not in the sense that its statements describe what really exists. Only
in the sense that they are valid for anybody. This I express by saying
they are but ``weakly objective''. In a way, the attempts at building up
theories of measurement that were made after Bohr's time may be viewed
as efforts aimed at imparting to ``orthodox'' quantum mechanics the
status of a strongly objective theory, but it can be considered that
these efforts failed. \par

Hence, quantum mechanics as we know it is not
ontologically interpretable. This is not necessarily to be considered
as a defect but it implies that, in the realm of interpretational
problems such as the one here on hand, we should not argue as if it
were. In particular the ``collapse riddle'' should prevent us from
tacitly assuming that the wave function possesses in every
circumstances all the attributes of reality. In fact the safest way to
make use of the wave function is just to consider it as a component of
a computational algorithm (or ``rule'') that enables us to know the
probability we have of observing such and such a measurement outcome on
a system $S$ when we know how $S$ was prepared. Such measurement outcomes,
described in a kind of a realist {\it language} (the pointer is at such and
such a place etc.) may then be considered as elements of an empirical
reality constituted by the phenomena understood in a Kantian sense,
that is, as more than mere appearances but less than elements of some
ontologically defined Reality, since they depend partly on us. \\

\newpage
\noindent {\bf 4. The proper mixture cannot be created.} \\

 This is the title of one of
Kirpatrick's sections and, in a sense, the statement it conveys -
presented by its author as an objection to my views - is a correct one.
Indeed, if we kling to the (here undisputed) view that a pure quantum
state $|\Psi >$ of a system yields the maximal information that can be
obtained on this system we have to consider that the idea of an
observable physical diversity that would exist, independently of us, in
an ensemble of isolated systems $D$ described by $|\Psi >$ is
self-contradictory, and that the time evolution operator cannot all by
itself generate that physical diversity. Concerning the case in which
the systems $D$ are composite we must also admit that choosing to focus
our attention on such and such features of $|\Psi >$ (by mathematical
operations such as taking partial traces and so on) will never make $|\Psi >$
generate that physical diversity, which, however, constitutes the
defining characteristic of what von Neumann called mixtures and I
called proper mixtures.  So, in this Kirkpatrick is right. Moreover, I
showed ([3], chapter 17) that, in this respect, replacing such a pure
state $|\Psi >$ by a mixture is of no help. \par

However, this means that the
diversity in question has to be rejected, and that the same is
therefore true concerning the ignorance interpretation of the
``improper'' mixtures (of subsystems of the $D$'s). But still  we do
observe diversity when, on statistical ensembles of systems, we perform
observations. So, we face a difficulty. To study it let us consider the
way Kirkpatrick presented his ``proof'' (that ``the proper mixture cannot
be created''). We must here quote him at some lengh. He asked ``How might
we go about creating a mixture, in particular a proper mixture?''. And
he continued: ``We return to von Neumann's original description of the
mixed state (echoed by d'Espagnat for the case of the proper mixture).
The preparation of the system $S$ varies randomly among the possible
output states $\{|\alpha_j>\}$; when $S$ is prepared in the state $|\alpha_j>$, the state
of its relevant environment ${\cal E}$ (a system external to $S$ such that $S+{\cal E}$ has
no correlations with its exterior) is $|\eta_j>$ and the composite system is
described by the state $|\alpha_j\eta_j>$. Because $S+{\cal E}$ has no exterior
correlations, these states are indistinguishable; the
Indistinguishability Rule requires the state of $S+{\cal E}$ to be pure, the sum
$|\Psi^{S+{\cal E}}> = \Sigma_s \gamma_s|\alpha_s\eta_s>$''. He then rewrote $|\Psi^{S+{\cal E}}>$ in the form of a
bi-orthogonal Schmidt-like decomposition, and claimed that the state of
$S$ is the improper mixture obtained by tracing out ${\cal E}$ on $|\Psi^{S+{\cal E}}><\Psi^{S+{\cal E}}|$.
His conclusion was: ``it is not possible to create d'Espagnat's proper
mixture''. \par

I already explained why I consider that, in contexts of this
type, Kirkpatrick's use of the Indistinguishability Rule is faulty.
Here, however, this is by no means an essential point for, in this
passage, Kirkpatrick explicitly considered the preparation of a
mixture, taking the environment of the system into account. Now, when
we think in terms of preparation we may without generality loss (see
above) imagine that the combined $S+ {\cal E}$ system on which the mixture of the
$S$'s is prepared is a pure state. \par

In fact, the trouble with
Kirkpatrick's approach lies at a much deeper level, tightly connected
with the fact, commented on in the foregoing section, that ``orthodox''
quantum mechanics is not (as Bohm used to say) ontologically
interpretable; that, in other words, the Reality it describes is but
Empirical Reality. To see what the said trouble consists of, it
suffices to have a careful look at the structure of Kirkpatrick's above
reported argument. What is crucial in it is the specification that the
considered environment, ${\cal E}$, of $S$ be ``[a system external to $S$] such that
$S+{\cal E}$ has no correlations with its exterior''. When (or assuming that)
this is the case, everything that Kirkpatrick wrote nicely follows... but
on the other hand it is just pure mathematics, deprived of any bearing
on possible observations since, in order that we should be be able to
observe anything, some interaction must take place between us and
either $S$ or ${\cal E}$ or both. Admittedly it could be assumed that the
observing elements are themselves parts of ${\cal E}$. But then we would have to
face a dilemma: either we consider that these ``observing elements'' are
inanimate objects such as counters -- but then nothing is gained since
these counters must themselves be observed from outside ${\cal E}$ -- or we
assume that we, the ``observers'', are ourselves within ${\cal E}$. However, we
then have to face a riddle that all the (numerous) attempts at building
up a consistent quantum measurement theory have not been able to
resolve, namely the ``and-or'' enigma: wherefrom does it come that we
have the feeling -- nay the ``certainty''! -- of being either in the $|v_1>$
state or in the $|v_2>$ state even though the finest possible description
of the whole state of affairs is of the type $c_1|u_1>|v_1> + c_2|u_2>|v_2>$
i.e. contains both $|v_1>$ and $|v_2>$~? Note that, obviously, in this matter
dropping Kirkpatrick's condition that $S+{\cal E}$ should have no correlations
with its exterior would not help. \par

And yet, since we are dealing not
with just pure mathematics but with physics, we simply cannot do as if
the just mentioned feeling -- or rather, certainty -- we have of always
seeing pointers and other objects at definite places did not exist. One
way of resolving this paradox (the only way I know of!) is to keep in
mind something like Bohr's above quoted assertion, that is, consider
that the purpose of physics is not to describe ``Reality as it really
is'' but, less ambitiously, to synthetically describe our communicable
experience. Within the framework of such an approach wave functions,
state operators and so on are, to repeat, essentially means of
prediction of observations and the elements of the ensemble that we
work with are pieces of empirical reality. And indeed, one of the most
remarkable facts that quantum mechanics revealed is that, far from
being restricted to the description of some -- conjectured --
man-independent Reality, mathematics are fully suitable for describing
such a man-dependent empirical reality. \par

In particular, mathematics
yield convenient tools for synthetizing our predictive knowledge
concerning systems on which we assume that measurement have somehow
been done without their outcomes being known to us (or, more generally,
of which we assume that they interacted with macroscopic objects that
can be treated classically). Von Neumann's ``ignorance'' mixtures (my
``proper mixtures'') are precisely the tools in question. And, within the
empirical reality approach that we are here considering, to say that
these tools cannot be created is no more true. Take a beam of spin 1/2
particles polarized along $Ox$. Send it through an inhomogeneous magnetic
field directed along $Oz$ and put counters on the two emerging paths. If,
in accordance with what the ``man in the street'' would say, you claim
that, corresponding to each one of the beam particles,  one (only) of
the two counters did really click (in the empirical reality sense),
then you have to grant that (in the same sense), beyond the counters
the ensemble of the particles is a proper mixture. \par

It is true that, in
principle, you are not quite obliged to take this standpoint. You may
boldly say: ``to claim that the counters either really clicked or did
not really click is an overnaive conception of what Reality is'', and
then you can resort to Kirkpatrick's reasoning and state that the
mixture is an improper one. If, notwithstanding the conceptual
difficulties, you kling to the view that the wave function is an
element of ``Reality as it really is'', then -- if, moreover, you believe
quantum mechanics is universal -- you are even forced to take up the
latter viewpoint. But, to repeat, the conceptual difficulties just
alluded to are, in fact, insuperable. The Bohr-like view that quantum
mechanics is really a weakly objective theory is therefore a
considerably more reasonable (and I would even say: ``scientific'')
approach. And, to repeat, within its realm the notion of proper
mixtures is fully valid. \\

\noindent {\bf 5. Conclusion.} \\

 I think that, in substance, I answered all of
Kirkpatrick's objections, including those bearing explicitely on
measurement. In the latter, what Kirkpatrick stressed is, in fact, just
that, when a measurement occurs, the values of the measured observable
must necessarily be correlated with something outside the system. This
of course is true but I have already considered the matter in Section 4
above since the ``something outside the system'' is obviously a part of
${\cal E}$. As his parenthesis beginning by ``Curiously...'' reveals, what
Kirkpatrick did not realize is that, in the case of a measurement, the
same subensemble of measured systems should be considered either as a
proper or as an improper mixture, according to whether we choose to
consider the instruments as being ``on the classical side'' or ``on the
quantum side'', that is, according to where we decide to situate -- by
thought -- the quantum-classical cut (which, according to the views here
reported in Section 3, separates the domain in which we can use a
realist language from the one in which we cannot, and depends on what
we are interested in). \par

One last but (I hope) not very significant point
concerns Kirkpatrick's remark, at the end of his Section 5, relative to
the ignorance interpretation. He there speaks of the ``temptation'' to
the interpretation of mixtures by ignorance, and he claims that the
fact all mixtures are improper gives a clearer understanding of the
said ``temptation''. This language seems to mean that an ignorance
interpretation of the improper mixture is actually inconsistent. But
then, when, in the first section of his paper, we read: ``D'Espagnat
claims that an ignorance interpretation of the improper mixture is
inconsistent'' we are somewhat at a loss. We get to wonder where,
according to him, the difference between us lies. Still, since he
repeateadly speaks of my ``error'' and emphasizes its importance, there
must exist some difference! One conjecture would be  that he situates
``diversity'', ``well-defined values'' and, therefore, ``ignorance'' at the
extreme end of the von Neumann chain: within some nonphysical ``taking
cognizance of definite values'' element in it. On the other hand, it is
clear that in his analysis of Hugues' argument he took positions quite
incompatible with this hypothesis. I, therefore, have no clue. In last
resort I cannot completely rule out the hypothesis that, on this point,
Kirkpatrick's approach was not entirely consistent.\\

\noindent {\bf References} \\

\noindent  1. K.A. Kirkpatrick, {\it arXiv:quant-ph/0109146 v2 21 Oct 2001}. \\ 

\noindent 2.  B. d'Espagnat, {\it Conceptions de la physique conremporaine}, Hermann, Paris
1965; contribution to {\it Preludes in Theoretical Physics:In Honour of
V.F.Weisskopf}, A.De Shalit, H.Feshbach and L.Van Hove Eds.,
Noth-Holland, Amsterdam, 1966, p. 185. \\ 

\noindent 3. B. d'Espagnat, {\it Conceptual
Foudations of Quantum Mechanics, Second Edition}, Addison-Wesley,
Reading Mass. 1976, 4d. ed. Perseus Books 1999. \\ 

\noindent 4. B. d'Espagnat, {\it Veiled
Reality}, Addison-Wesley, Reading Mass. 1995; in French: {\it Le r\'eel voil\'e},
Paris, Fayard, 1994. \\ 

\noindent 5. J.von Neumann, {\it Mathematical Foundations of
Quantum Mechanics}, Princeton U. Press Princeton N.J. \\

\noindent 6. R.I.G. Hughes,
{\it The Structure and Interpretation of Quantum Mechanics}, Harvard
University Press, Cambridge MA., 1989. \\

\noindent 7. N. Bohr, {\it Quantum Physics and
Philosophy -- Causality and Complementarity, contribution to Philosophy
in the Mid-Century}, R.Klibansky Ed., La Nuova Italia Editrice, Florence
1958; reprinted in: Essays 1958-1962 on Atomic Physics and Human
Knowledge, A.Bohr Ed., printed by Richard Clay and Co., Bungay,
Suffolk, 1963.

\end{document}